\documentclass[12pt,a4paper]{article}
\usepackage{url}
\makeatletter
\newcounter{lineno}
\def\verbatimlisting#1{\setcounter{lineno}{0}%
    \begingroup{\footnotesize} \@verbatim \frenchspacing \@vobeyspaces
\parindent=20pt
    \everypar{\stepcounter{lineno}\llap{\thelineno\ \ }}\input#1
    \endgroup
}
\makeatother

\newcommand{\reduce}{{\sc reduce}}

\newcommand{\Ord}[1]{{\cal O}\left(#1\right)}

\newcommand{\cR}{{\cal R}}

\renewcommand{\vec}[1]{{\bf #1}}
\newcommand{\vel}{{\vec q}}

\newcommand{\grad}{\nabla}
\newcommand{\divv}{\nabla\cdot}

\begin{document}
\title{\sf Proper initial conditions for the lubrication model of the
flow of a thin film of fluid }
\author{S.A.~Suslov and A.J.~Roberts\thanks{Dept.\ Mathematics \&
Computing, University of Southern Queensland, Toowoomba, Qld~4352,
Australia.  E-mail: \texttt{ssuslov@usq.edu.au} and
\texttt{aroberts@usq.edu.au} respectively.}}
\date{5 April 1998}

\maketitle

\begin{abstract}
A lubrication model describes the dynamics of a thin layer of fluid
spreading over a solid substrate.
But to make forecasts we need to supply correct initial conditions to
the model.
Remarkably, the initial fluid thickness is \emph{not} the correct
initial thickness for the lubrication model.
Theory recently developed in \cite{Roberts89b,Roberts97b} provides the
correct projection of initial conditions onto a model of a dynamical
system.
The correct projection is determined by requiring that the model's
solution exponentially quickly approaches that of the actual fluid
dynamics.
For lubrication we show that although the initial free surface shape
contributes the most to the model's initial conditions, the
initial velocity field is also an influence.
The projection also gives a rationale for incorporating miscellaneous
small forcing effects into the lubrication model; gravitational
forcing is given as one example.
\end{abstract}


\tableofcontents

\section{Introduction}

The flows of thin films of fluids are encountered in many engineering
and biological applications.
They include: the flow of rainwater on a road or windscreen or other
draining problems \cite{Chang94}; paint and coating flows
\cite{Ruschak85,Tuck90a}; the flow of many protective biological
fluids \cite{Grotberg94}; and other coating, painting and drying
processes \cite[e.g.]{Moriarty91,Kalliadasis94,Wilson95}.
The fluid film thickness and the average fluid flux are the main
characteristics of interest in these applications.
The fine details of the actual local velocity and pressure fields
typically are of less practical importance.
For this reason the various approximations have been constructed over
the past decades
\cite[e.g.]{Chang94,Prokopiou91b,Tuck90a,Roberts94c,Jensen97} to model
the evolution of the fluid flow in various geometries \cite{Roy96} and
parameter regimes \cite{Frenkel97}.
We consider herein the basic nondimensional lubrication model for
surface tension dominated flows,
\begin{equation}
        \eta_t \approx
        -\frac{1}{3}\partial_x\left(\eta^3\eta_{xxx}\right)\,,
\label{hm}
\end{equation}
where $\eta(x,t)$ is the thickness of the fluid film spreading over a
solid substrate (at $y=0$).
Centre manifold theory \cite{Carr81} provides a generic and systematic
procedure for deriving such models for a wide variety of fluid flows.
Recently, Roberts \emph{et al.}~\cite{Roberts96a,Roy96} showed how this well
established lubrication model of thin film flow is rigorously derived
from the governing Navier-Stokes equations (outlined in
Section~\ref{Sflow}) using a computer algebra implementation of centre
manifold techniques.
The model is derived under the assumption that longitudinal derivatives,
$\partial/\partial x$, are small---the slowly varying assumption---as
used extensively in creating models of shear dispersion,
\cite[e.g.]{Mercer94a,Watt94c}.
The centre manifold based algorithm provides a straightforward
derivation of the model up to arbitrarily high order
\cite{Roberts96a}, but in this work we limit ourselves to
consideration of the above leading order model.

It is frequently believed that the initial conditions for the models
of long-term evolution are not highly important because the asymptotic
state does not depend on a character of transient processes in the
system.
It was shown in \cite{Roberts89b,Cox93b} that initial conditions {\em
can have} a long-lasting influence on forecasts.
This seems especially true for models of spatio-temporal dynamics such
as the lubrication model.
Similarly, in the case of dispersion in a channel \cite{Roberts89b} or
pipe \cite{Mercer94a} the long term location and spread of a pollutant
does depend upon the details of the initial release of the pollutant.
Thus the initial conditions for a model must be chosen carefully to
ensure the long term fidelity between the model and the physical flow.
It will be shown in Section~\ref{Sic} that the initial form of the
free surface for the lubrication model should be different
from the initial thickness of the physical fluid.
In particular, if the fluid initially has thickness $\eta_0(x)$ but
zero velocity and pressure then the lubrication model~(\ref{hm})
should be solved with initial condition that $\eta(x,0)=h_0(x)$ where
\begin{equation}
	h_0\approx \eta_0+\eta_0\eta_{0xx}\,.
	\label{h0simp}
\end{equation}
In general, the initial condition $h_0$ for the model is the
non-trivial function of initial velocities and pressure distributions
given by~(\ref{h_0})--(\ref{h0n}).
The argument for these initial conditions is based upon the dynamics
near the low-dimensional centre manifold, that is, upon the physics of
the approach to the lubrication model.
The general arguments, developed in \cite{Roberts89b} and recently
refined in \cite{Roberts97b}, are based upon the geometric picture
provided by centre manifold theory.
The principle aim of this paper is to apply this general framework to
the considerable complications of the infinite dimensional dynamics of
thin film fluid flow and so derive~(\ref{h0simp}) and its
generalisations.  This is the first time that correct initial conditions
have been obtained for lubrication theory.

An interesting aspect of the general analysis developed in
\cite{Roberts89b} and \cite{Roberts97b} is that the projection of
initial conditions also gives a rationale for treating small forcing
of the dynamics.
This connection was more fully explored by Cox \& Roberts \cite{Cox91}
who discussed the effects upon the centre manifold and the evolution
thereon for time dependent forcing.
In Section~\ref{Sgrav} we apply the projection to a gravitational
forcing of the thin fluid layer to verify the veracity of the classic
model
\begin{equation}
	\eta_t\approx -\frac{1}{3}\partial_x\left[ \eta^3\left(
	\eta_{xxx} +B\sin\theta -B\cos\theta \,\eta_x  \right) \right]\,,
	\label{Elubg}
\end{equation}
where $B$ is the nondimensional magnitude of gravity and $\theta$ is
the downwards angle of the substrate.
This is derived here as just one example of a very general result that
applies to all small forcing effects upon the fluid flow.

\section{The lubrication model of fluid flow}
\label{Sflow}

We consider a two-dimensional flow of thin film of Newtonian fluid along a
flat horizontal substrate. The free surface is given by $y=\eta(x,t)$, where
$x$ and $y$ are horizontal and vertical coordinates respectively. The flow,
with velocity $\vel=(u,v)$ and pressure $p$, is governed by the incompressible
Navier-Stokes equations
\begin{eqnarray}
        \vel_t+\vel\cdot\grad\vel
        & = & -\frac{1}{\rho}\grad p+\nu\grad^2\vel\,,
\end{eqnarray}
supplemented by the continuity equation
\begin{eqnarray}
        \divv\vel & = & 0\,,
\label {cont}
\end{eqnarray}
non-slip boundary conditions on the bottom
\begin{equation}
        \vel=\vec 0\quad\mbox{on $y=0$ ,}
        \label{svebbc}
\end{equation}
and tangential stress and normal stress conditions on the free surface
\begin{eqnarray}
        2\eta_{x}(v_y-u_x)+\left(1-\eta_{x}^2\right)(u_y+v_x) =0
        \quad\mbox{on $y=\eta$ ,}
        \label{svetc}
        &&\\
        \left(1+\eta_{x}^2\right)p
        = 2\mu\left[v_y+\eta_{x}^2u_x
        -\eta_{x}\left(u_y+v_x\right)\right]   &&\nonumber\\
        {} -\frac{\sigma\eta_{xx}}{\sqrt{1+\eta_{x}^2}}
\quad\mbox{on $y=\eta$ ,}&&
\end{eqnarray}
respectively, as discussed in detail by Roberts \cite{Roberts96a}. We close the
problem with the kinematic condition relating the velocity of the fluid on the
surface to the evolution of the free surface:
\begin{equation}
     \eta_t = v-u\eta_x\quad\mbox{on $y=\eta$ .}
        \label{svekc}
\end{equation}
In the above equations $\rho$ is the fluid density, $\nu$ is the kinematic
viscosity, and $\sigma$ is the coefficient of the surface tension. The fluid
film is assumed to be so thin that the gravity force in the momentum
equations can be neglected (see the discussion in \cite{Roberts96a}) at least
initially.

We non-dimensionalise the governing equations using a typical film thickness
$H$ as a reference length, reference time $\mu H/\sigma$ (where
$\mu$ is the dynamic viscosity of the fluid), reference speed $\sigma/\mu$,
and reference pressure $\sigma/H$. On this small scale, fluid viscosity
is strong and the fluid layer is of very large extent laterally.
The non-dimensional Navier-Stokes and continuity equations then become
\begin{eqnarray}
        \cR\left(\vel_t+\vel\cdot\grad \vel\right) & = &
        -\grad p+\grad^2\vel \,,
        \label{svefq} \\
        \divv\vel & = & 0\,,
        \label{svefc}
\end{eqnarray}
where $\cR=\rho \sigma H/\mu^2$ is a Reynolds number.
They are complemented by the normal stress condition
\begin{equation}
        \left(1+\eta_{x}^2\right)p=
        2\left[v_y+\eta_{x}^2u_x-\eta_{x}\left(u_y+v_x\right)\right]
  -\frac{\eta_{xx}}{\sqrt{1+\eta_{x}^2}}
\quad\mbox{on $y=\eta$ ,}
        \label{svenc}
\end{equation}
and equations~(\ref{svebbc}), (\ref{svetc}) and~(\ref{svekc}) which remain
symbolically unchanged under the nondimensionalisation.

In such a thin layer of fluid, the infinite number of horizontal shear
modes decay exponentially quickly through viscous dissipation acting
across the thin film.
Thus in the long term, the dynamics are driven by surface tension
trying to flatten surface curvature.
Centre manifold theory \cite{Carr81} is used in such circumstances to
systematically derive the low-dimensional model of the long term
evolution, here the lubrication model~(\ref{hm}) for the fluid layer's
thickness $\eta$, see~\cite[\S3]{Roberts96a} for more introductory
detail.
The approximate form of the lubrication model for such a flow is
obtained as a formal expansion in orders of the $x$-derivatives under
the assumption that these derivatives are small.
Although the model can be developed to an arbitrary order of spatial
derivatives using the iterative computer algebra algorithm suggested
in \cite{Roberts96a}, the expressions for higher order approximations
are very involved and thus we present here only the lowest order model.
To errors of fifth-order in $\partial_x$ and parameterized by the
free surface thickness $\eta$, the centre manifold $\vec
v=(u(\eta),v(\eta),p(\eta),\eta)$ is given by
\begin{eqnarray}
        u & \approx & \left(y\eta-\frac{1}{2}y^2\right)\eta_{xxx}\,,
\label{um}\\
        v & \approx & -\frac{1}{2}y^2\eta_{x}\eta_{xxx}
        +\left(\frac{1}{6}y^3-\frac{1}{2}y^2\eta\right)\partial_x^4\eta\,,  \\
        p & \approx & -\eta_{xx}
+\frac{3}{2}\eta_{x}^2\eta_{xx}-\left(\eta+y\right)\eta_{x}\eta_{xxx}
         -\left(\frac{1}{2}\eta^2+y\eta-\frac{1}{2}y^2\right)\partial_x^4\eta\,,
         \label{pm}
\end{eqnarray}
where $\eta$ evolves according to~(\ref{hm}).
Observe that up to this order the model does not depend on the Reynolds
number---fluid inertia is negligible.
The lubrication model~(\ref{hm}) is the basic model for the dynamics
of thin fluid films.

\section{Project the initial conditions}
\label{Sic}

In order to use the lubrication model~(\ref{hm}) to make forecasts, it
should be supplemented with initial conditions.
Roberts \cite{Roberts89b} has shown that determining the correct
initial conditions is a nontrivial problem.
Remarkably, in general the initial value of $\eta$ for
model~(\ref{hm}) differs from the initial fluid thickness for the
physical problem~(\ref{svebbc})--(\ref{svenc}).
To distinguish between the two, we denote the initial fluid thickness
by $\eta_0$ and use $h_0$ to denote the initial conditions for
model~(\ref{hm}) of the fluid's evolution.
The main task of this paper is to show how to determine $h_0$ as a
function of the initial fluid state.

To define the proper initial conditions for model~(\ref{hm}) we follow
the procedure outlined by Roberts \cite{Roberts97b} and examine the
dynamics in the vicinity of the centre manifold.
We start by linearizing the governing equations about the centre
manifold~(\ref{um})--(\ref{pm}) by writing the fluid variables as the
sum $(u,v,p,\eta)+(u^\prime,v^\prime,p^\prime,\eta^\prime)$ where
primed quantities are the assumed small displacement from the centre
manifold, and so their products are neglected.
The resulting system is
\begin{eqnarray}
  \cR\left( \vel^\prime_t+\vel^\prime\cdot\grad \vel+ \vel\cdot\grad
    \vel^\prime\right)+\grad p^\prime-\grad^2\vel^\prime &
  = & 0 \,, \label{lvel}\\
  \divv\vel^\prime & = & 0\,,
\end{eqnarray}
with the boundary conditions at $y=\eta$
\begin{equation}
  \eta^\prime_t-v^\prime+u^\prime\eta_x+u\eta^\prime_x
+u_y \eta_x \eta^\prime-v_y \eta^\prime = 0 \,,
  \label{lkin}
\end{equation}
\begin{eqnarray}
  2\eta_{x}^\prime\left(v_y-u_x\right)+
  2\eta_{x}\left(v^\prime_y-u^\prime_x\right)
  +\left(1-\eta_{x}^2\right)\left(u^\prime_y+v^\prime_x\right)
  -2\eta_{x}\eta_x^\prime\left(u_y+v_x\right) \nonumber\\
  +2\eta_x(v_{yy}-u_{xy})\eta^\prime
  +\left(1-\eta_x^2\right)(u_{yy}+v_{xy})\eta^\prime=0 \,,
\end{eqnarray}
\begin{eqnarray}
  \left(1+\eta_{x}^2\right)p^\prime+2\eta_x\eta_x^\prime p+
  \left(1+\eta_x^2\right)p_y\eta^\prime
   +\frac{\eta_{xx}^\prime}{\sqrt{1+\eta_{x}^2}}
  -\frac{\eta_{xx}\eta_x\eta_x^\prime}{(1+\eta_{x}^2)^{3/2}} =  \nonumber\\
  2\left[v_y^\prime+\eta_{x}^2u_x^\prime+2\eta_x\eta_x^\prime u_x
  -\eta_{x}\left(u_y^\prime+v_x^\prime\right)
  -\eta_{x}^\prime\left(u_y+v_x\right)\right]\label{lbc}\\
  +2\left[v_{yy}+\eta_x^2u_{xy}-\eta_x(u_{yy}+v_{xy})\right]\eta^\prime\,,
  \nonumber
\end{eqnarray}
and the homogeneous boundary conditions for the velocity $\vec
q^\prime={\vec 0} $ at $y=0$.  The above equations describe the
dynamics of the fluid near the centre manifold~(\ref{um})--(\ref{pm}).

Typically, the initial conditions $\vec
u_0=(u_0(x,y),v_0(x,y),p_0(x,y),\eta_0(x))$ for the original fluid
layer equations~(\ref{svebbc})--(\ref{svenc}) do not belong to the
low-dimensional centre manifold $\vec v$ given
by~(\ref{um})--(\ref{pm}).
Thus they cannot be used directly as a starting point for
model~(\ref{hm}).
As shown in \cite{Roberts89b} and \cite{Roberts97b} the proper model
initial condition is the projection $\vec
v_0=(u(h_0),v(h_0),p(h_0),h_0)$ from $\vec u_0$ to the centre manifold
along the isochron---in the state space an isochron is a surface of
all the initial states which have the same long-term dynamics on
the centre manifold (up to an exponentially small error).
Consequently, the model initial conditions are determined to satisfy
\begin{equation}
  \langle\vec z,\vec u_0- \vec v_0 \rangle=0\,,
  \label{ic}
\end{equation}
where $\vec z=(u^\dag,v^\dag,p^\dag,\eta^\dag)$ is a vector orthogonal
to the direction of projection (the dagger is used to denote field
quantities in the adjoint space).  Here the inner product is defined
for four component vector fields
\begin{eqnarray*}
	 &  & {\vec a}=(a^1(x,y,t),a^2(x,y,t),a^3(x,y,t),a^4(x,t))
	 \quad\mbox{and}  \\
	 &  & {\vec b}=(b^1(x,y,t),b^2(x,y,t),b^3(x,y,t),b^4(x,t))
\end{eqnarray*}
as
\begin{equation}
  \langle {\vec a},{\vec b} \rangle \equiv
  \int_{-\infty}^\infty\int_0^\eta
  (a^1b^1+a^2b^2+a^3b^3)\,dy\,dx\,+
  \int_{-\infty}^\infty a^4b^4\,dx\,.
  \label{ip}
\end{equation}
According to the arguments developed in \cite{Roberts89b} and refined
in \cite{Roberts97b}, the defining vector of the projection, $\vec z$,
satisfies the dual equation
\begin{equation}
  {\cal D} \vec z =\langle \cal D \vec z, \vec e \rangle \vec z\,,
  \label{dual}
\end{equation}
where $\vec e$ is the local tangent vector to the
centre manifold
\begin{equation}
  \vec e= \frac{\partial}{\partial \eta} \left[\begin{array}{c}
      u\\v\\p\\\eta \end{array} \right]=
  \left[\begin{array}{c}
      0\\
      0\\
      -\partial^2_x\\
      1
    \end{array} \right]+\Ord{\partial_x^3}\,,
\label{ecm}
\end{equation}
and the dual operator $\cal D$ is obtained from equations adjoint
to~(\ref{lvel})--(\ref{lbc}) with respect to the inner product defined
as
\begin{equation}
  \langle\!\langle {\vec a},{\vec b} \rangle\!\rangle \equiv
  \int_0^t\langle {\vec a},{\vec b} \rangle\,d\tau\,.
\end{equation}
Higher order derivative terms in~(\ref{ecm}) can be easily computed
but, as will be shown later, the given second order truncation will
suffice for finding the initial conditions to the first few orders.
Note that the local tangent to the centre manifold is a vector
operator rather than just a vector function as occurs in the finite
dimensional cases discussed in \cite{Roberts97b}.  Being introduced
into the inner product~(\ref{ip}), it acts on the other vector
involved before the integration is performed.  Using the above inner
products, the adjoint expressions of~(\ref{lvel})--(\ref{lbc}) leading
to the dual operator ${\cal D}$ are:
\begin{equation}
{\cal D}\vec z=\left[\begin{array}{c}
{\cal R}\left(u_t^\dag-u^\dag u_x-v^\dag v_x+uu^\dag_x +vu^\dag_y \right)
+p^\dag_x+u^\dag_{xx}+u^\dag_{yy}\\
{\cal R}\left(v_t^\dag-u^\dag u_y-v^\dag v_y+uv^\dag_x +vv^\dag_y \right)
+p^\dag_y+v^\dag_{xx}+v^\dag_{yy}\\
u^\dag_x+v^\dag_y\\
\eta^\dag_t+\eta^\dag-p^\dag+v^\dag\eta_{xx}+u^\dag_x+2v^\dag_x\eta_x-v^\dag_y
\quad \mbox{on $y=\eta$}
\end{array}\right]\,.
\end{equation}
The adjoint velocities satisfy homogeneous boundary conditions
$\vel^\dag={\vec 0}$ at $y=0$.
The adjoint boundary conditions for the velocities at $y=\eta$ are
represented by expressions too long to be given here.
Instead in the Appendix we include computer algebra code written in
{\sc reduce} for obtaining them and the dual operator
${\cal D}$.
Periodic boundary conditions at $x=\pm \infty$ are used in the
derivation.

Given the above dual ${\cal D}$, system~(\ref{dual}) is solved
asymptotically assuming that it is possible to neglect higher order
derivatives with respect to $x$.
The treatment of $\partial_x$ as small is equivalent to the assumption
of slow variation in $x$.
The iterative algorithm is quite similar to the one described in
\cite{Roberts96a} and used to derive the centre manifold
model~(\ref{um})--(\ref{pm}) \& (\ref{hm}).
Thus here we just make a few notes on the specifics of the first few
iterations.
In essence the procedure is as follows: we start by solving the
equations neglecting all $x$ derivatives and then in further
iterations compute the corrections associated with these derivatives
of functions found at previous iterations.
Owing to the special form of the vector $\vec e$, the right-hand side
of~(\ref{dual}) remains zero during the first few iterations required
to obtain the leading order (in $\partial_x$) expressions for $\vec z$.
It is easier to first look for the solution $\vec z$ in the functional
\begin{eqnarray}
  \vec z= \left[\begin{array}{c}
      u^\dag\\v^\dag\\p^\dag\\\eta^\dag \end{array} \right]&\approx&
  \left[\begin{array}{c}
      0\\0\\\eta^\dag\\\eta^\dag
    \end{array} \right]+
  \left[\begin{array}{c}
      \eta_x^\dag\left(y\eta-\frac{1}{2}y^2\right)\\0\\0\\0
    \end{array} \right] \label{z}
    \\&&{}+
  \left[\begin{array}{c}
      0\\\eta_{xx}^\dag\left(\frac{1}{6}y^3
        -\frac{1}{2}y^2\eta\right)
      -\frac{1}{2}\eta_{x}^\dag\eta_xy^2\\
      \eta_{xx}^\dag\left(y\eta-\frac{1}{2}y^2\right)
      +\eta_{x}^\dag\eta_xy
      +\frac{1}{2}\partial_x\left(\eta_{x}^\dag\eta^2\right)
      \\0
    \end{array} \right]\,,\nonumber\\
  \mbox{s.t.}\quad \eta^\dag_t&\approx&0 \,.
\end{eqnarray}
From the structure of the successive corrections to the solution for $\vec z$
we deduce that the initial conditions for the model are influenced the most
by the initial form of the fluid surface. This is the expected result for such
a surface tension dominated flow. The initial horizontal velocity field has a
secondary effect on the flow (corresponding terms in~(\ref{z}) appear only in
the second iteration) primarily as a response to the horizontal pressure
gradient induced by the surface curvature. The vertical motion is even less
important since it is severely restricted by the small thickness of the fluid
layer.

An additional condition must be exploited to determine $\eta^\dag$ as
an asymptotic expansion in $\partial_x$.  It is done using the
normalisation condition $\langle\vec z,\vec e \rangle=1$ (see
\cite{Roberts97b}) which upon~(\ref{ecm}) leads to
\begin{equation}
  \int_{-\infty}^\infty\left(\eta^\dag
    -\int_0^{\eta}p^\dag_{xx}\,dy\right)dx+\Ord{\partial^3_x}=1\,.
  \label{norm}
\end{equation}
At the leading order we then obtain
\begin{equation}
  \int_{-\infty}^\infty\eta^\dag\,dx=1\,.
\end{equation}
This condition does not provide any unique solution for $\eta^\dag$ but
rather a continuum of linearly independent localised functions. Without any
loss of generality we choose the linearly independent solutions
\begin{equation}
  \eta^\dag(x;x_*)=\delta(x-x_*)+\Ord{\partial^2_x}\,.
  \label{hd0}
\end{equation}
where $x_*$ is an arbitrary point and $\delta$ is the Dirac delta
function.  At the next iteration we require the second order
contribution to~(\ref{norm}) to vanish.  This results in
\begin{equation}
  \eta^\dag(x;x_*)=\delta(x-x_*)+\delta^{\prime\prime}(x-x_*)\eta(x,t)
  +o\left(\partial^2_x\right)\,.
  \label{hd2}
\end{equation}
This expression for $\eta^\dag$ is used in~(\ref{z}) to determine
$\vec z$, the defining vector of the proper projection onto the centre
manifold.

Lastly, we use $\vec z$ in~(\ref{ic}) to project an initial condition.
Requiring the integrand in equation~(\ref{ic}) to vanish and
taking into account~(\ref{z}), (\ref{hd0}) and~(\ref{hd2}) we obtain
\begin {eqnarray}
  & \eta_0-h_0+\overline {p_0}+\partial_x
  \left[\overline {u_0y\left(\frac{y}{2}-h_0\right)}\right] & \nonumber \\
  &-\partial_x
  \left[h_{0x}\left(\overline {p_0(h_0+y)
        -\frac{1}{2}v_0y^2}\right)\right] +
  h_0 h_{0xx}& \\
  & +\partial_x^2\left[ h_0\eta_0-h_0^2+
    \overline{p_0\left(\frac{h_0^2-y^2}{2}+h_0(y+1)\right)
      -\frac{v_0}{2}\left(h_0y^2-\frac{y^3}{3}\right)}\right]
    &\approx 0\,,\nonumber
\end{eqnarray}
where the notation $\overline f\equiv\int_0^{\eta_0}f\,dy$ is
introduced.  This equation determines $h_0$ and can be solved
iteratively as well.  The first three iterations produce
\begin{equation}
  h_0\approx h_{00}+h_{01}+h_{02}\,,
  \label{h_0}
\end{equation}
where
\begin{eqnarray}
  h_{00} &=&\eta_0+\overline{p_0}\,,\nonumber \\
  h_{01} &=&\partial_x
  \left[\overline{u_0y\left(\frac{y}{2}
        -h_{00}\right)}\right]\,,\label{h0n}\\
  h_{02} &=&
  -\partial_x\left[h_{01}\overline{u_0y}\right]
  +h_{00}h_{00xx}-\partial_x
  \left[h_{00x}\left(\overline{p_0(h_{00}+y)
        -\frac{1}{2}v_0y^2}\right)\right] \nonumber\\
  & &+\partial_x^2\left[\overline{
      p_0\left(\frac{h_{00}^2-y^2}{2}+h_{00}y\right)
      -\frac{v_0}{2}\left(h_{00}y^2-\frac{y^3}{3}\right)
      }\right]\,.\nonumber
\end{eqnarray}
Note some specific cases of interest.
\begin{itemize}

\item Parallel shear flow $v_0=p_0=0$, $\eta_0=1$ and $u_0=u_0(y)$.
According to the linear stability analysis, due to viscous dissipation
such a flow approaches the motionless state exponentially quickly.
Thus the centre manifold model, which disregards exponentially fast
transients, must give rise to a stationary solution.  In this case the
initial conditions~(\ref{h_0}) for the model and the model solution
itself are just $\vec v_0=\vec v(t)=(0,0,0,1)$ and do indeed
correspond to the motionless uniform fluid film.

\item Initially stationary fluid layer ($u_0=v_0=0$) with uniform
pressure equal to the atmospheric one ($p_0=0$) and curved free
surface $\eta_0\neq \mbox{const}$.
In this case $h_0\approx\eta_0(1+\eta_{0xx})$ and the initial
conditions for the model coincide with the initial film thickness only
to leading order.
Higher order terms tend to smooth out the initial distribution of the
model film thickness flattening ``hills'' and ``valleys''.
This can be interpreted in the following way.
The physical fluid which is initially motionless requires time for
acceleration i.e.~time to approach the centre manifold
(\ref{um})--(\ref{pm}) in which velocities are generally non-zero.
Dissipation during this transient acceleration leads to a decrease in
the energy of the system.
Since the energy of the system up to leading order in $\partial_x$ is
just the potential energy associated with the surface tension and is
proportional to the surface curvature the initial condition models the
energy loss by levelling out the free surface in comparison with the
original distribution.

\item Nonzero initial average pressure.
It leads to a change in the initial model fluid film thickness when
compared with that of the original problem.
In particular, locally positive initial pressure corresponds to
locally thicker fluid film in the model.
This is intuitively expected since the increased pressure inside the
fluid layer (imagine an underwater explosion) acts against the local
surface tension and leads to the appearance of the local ``hill'' on
the film surface.

\end{itemize}

Finally we note that the initial condition for the model is most
sensitive to the fluid film thickness and the local pressure whereas
the initial velocity field has just secondary effect on the long term
film dynamics.
This is not a surprise since, as noted in \cite{Roberts96a}, the
considered flow is essentially the creeping one and inertia effects
are less important than the influence of the surface tension or,
equivalently, of the surface curvature.

\section{Gravitational forcing as an afterthought}
\label{Sgrav}

In the previous sections the influence of gravity on the thin film
flow is neglected.  Here we demonstrate how such a forcing may be
added into the model using the projection derived for initial
conditions as argued in general in \cite{Roberts97b}.  The general
technique may be used to include physical processes into the
lubrication model \emph{after} developing the model.  The result given
here for gravity is just one specific example.

The correction to the model accounting for the gravity can be obtained
by iterative solution of the Navier-Stokes equations~(\ref{svefq})
where terms ${\vec f}=B(g_1,g_2)$ responsible for gravity are 
introduced in the right-hand side \cite[e.g.]{Roy96}.
Here $B=\rho |{\vec g}| H^2/\sigma$ (assumed finite but small) is the
Bond number \cite{Roberts96a} and $g_1$ and $g_2$ are components of the
non-dimensional gravity vector in $x$- and $y$-directions,
respectively.
Alternatively, considering gravity as a specific example of a forcing
which by arguments in \cite{Roberts97b} can be directly projected onto
the model~(\ref{hm}).
Geometrically, the centre manifold obtained for the dynamics without
forcing is deformed slightly when forcing is applied such that each
point of the unforced centre manifold is shifted along the isochron
passing through the original location of this point as discussed by
Cox \& Roberts \cite{Cox91}.
According to \cite{Roberts97b} the dynamics of the free surface
subject to forcing is described by the modified model~(\ref{hm})
\begin{eqnarray}
    \eta_t & \approx & -\frac{1}{3}\partial_x\left(\eta^3\eta_{xxx}\right)+q\,,
\end{eqnarray}
where $q$ is the projection of the forcing $\vec f$ of the fluid, namely
\begin{equation}
  q=\langle{\vec z},{\vec f} \rangle\,.
  \label{q}
\end{equation}
Typically, gravity is uniform in thin film flow applications and then
$(g_1,g_2)=(\sin \theta, -\cos \theta)$, where $\theta$ is the
downwards angle between a flat substrate and the horizontal.
Then upon using~(\ref{z}) and~(\ref{hd0}), (\ref{q}) immediately leads
to
\begin{eqnarray}
  q&=&B\int_{-\infty}^\infty\int_0^\eta \left\{g_1\delta^\prime_x\left(\eta
  -\frac{1}{2}y\right)y\right.\nonumber\\
& &+\left.\frac{1}{2}g_2 \left[\delta^{\prime\prime}_{xx}
\left(\frac{1}{3}y-\eta\right)-\delta^\prime_x\eta_x\right]y^2\right\}
\,dy\,dx \\
&=&-B\left[g_1\eta^2\eta_x+\frac{1}{3}g_2\partial_x
\left(\eta^3\eta_x\right)\right]\,,\nonumber
\end{eqnarray}
which is identical to the correction obtained from directly modelling
the forced equations~(\ref{svefq}) through assuming small Bond number
\cite{Roy96}.
The theory of initial conditions recently refined in \cite{Roberts97b}
also provides an elegant way of modifying the model to incorporate
forcing.

\section{Conclusions}

The proper initial conditions for the lubrication model of flow of
thin film is derived using the projection of the initial conditions
for the original problem onto the centre manifold representing the
lubrication model.
The obtained results are easily generalised to the case of isotropic
three-dimensional thin film flow.
Then the two-dimensional lubrication model
\begin{equation}
\eta_t\approx-\frac{1}{3} \divv
   \left(\eta^3\nabla^3\eta\right)
\end{equation}
should be solved with initial conditions given up to the first order by
\begin{equation}
h_0\approx\eta_0+\overline{p_0}
+\divv \overline{\vel_{0}y\left(\frac{1}{2}y-
\eta_0-\overline{p_0}\right)}\,.
\end{equation}
Here $\vel_{0}=(u_0,w_0)$ is the initial horizontal velocity field for
the original problem and $\nabla$ is a two-dimensional operator in
$xz$-plane.

\paragraph{Acknowledgement}
This work was supported by a grant from the Australian Research Council.

\appendix

\section{Computer algebra derives the dual}

\textsc{reduce}\footnote{At the time of writing, information about
\texttt{reduce} was available from Anthony C.~Hearn, RAND, Santa
Monica, CA~90407-2138, USA. E-mail: \tt reduce@rand.org} code for
determining the dual operator ${\cal D}$ along with its associated
boundary conditions:
\verbatimlisting{madj.red}

Below we comment on the listed {\reduce} program:

\begin{enumerate}
\item  {\sf Preliminaries.}
  \begin{itemize}

  \item $\ell$ 10--12 describes states slightly
    displaced from the centre manifold.

  \item $\ell$ 13--20 defines short hands and their properties for the
  expressions entering the free surface boundary conditions and
  appropriate algebraic and differential rules.

  \item $\ell$ 22--33 expresses the physical fluid equations and their
  boundary conditions.
  \end{itemize}

\item  {\sf The linearisation about the centre manifold.}

  \begin{itemize}

	  \item $\ell$ 35 makes use of the continuity equation to get rid
	  of $v_y$ and affirms that the free surface form $\eta^\prime$
	  does not depend on the vertical coordinate $y$.

	\item $\ell$ 37, 38 state that $v_y^\prime=-u_x^\prime$ in order
	to simplify the boundary conditions.  This definition has to be
	local to allow for the derivation of the adjoint continuity
	equation later.  Thus $\ell$ 41 clears this rule after it is used
	here.

	\item $\ell$ 36, 39 and 40 extracts the linearized kinematic
	and boundary conditions taking into account the variation in
	$y$ of the free surface itself.

	\item $\ell$ 44 extracts the linearized momentum and
	continuity equations.
    \end{itemize}

\item {\sf Determination of the adjoint equations.}
  \begin{itemize}

	\item $\ell$ 47 introduces the operator {\sf iii} which obtains
	the adjoints to the differential sub-operators entering linearized
	equations~(\ref{lvel})--(\ref{lbc}) through the integration by
	parts rules listed in $\ell$ 49--60.  Note that the volumetric
	integrals in~(\ref{ip}) after integration by parts contribute to
	the surface integral because the adjoint functions and their
	derivatives generally are not zero on the free surface which is a
	function of $x$ and $t$.  In addition, note that the rule in
	$\ell$ 53--54 uses the rule previously defined in $\ell$ 42.

  \item $\ell$ 61--62 forms the inner product~(\ref{ip}) taken with a
    negative sign for further convenience.

  \item Finally, $\ell$ 64--66 extracts the adjoint equations
  which result directly in the expression for the dual to be
  solved to obtain the model initial condition generating
  functions.
  \end{itemize}

\item {\sf Determination of the adjoint kinematic and boundary conditions.}

  This is done in three steps:
  \begin{itemize}

	\item Firstly, the $y$ derivatives of the unknown
	functions must be eliminated from the expressions for
	the adjoint boundary conditions since they remain
	undefined under the surface integration.  This is done
	by making use of the continuity equation to eliminate
	$v_y^\prime$ ($\ell$ 68--76) and the tangential stress
	boundary condition ($\ell$ 73 with so far not defined
	coefficients $f_i$) to eliminate $u_y^\prime$.
	Secondly, $p^\prime(x,\eta+\eta^\prime,t)$ is eliminated
	through the normal stress boundary condition ($\ell$
	74--75).

	\item The surface operator {\sf ii} is introduced in
	$\ell$ 77, which specifies the integration by parts
	rules ($\ell$ 79--83) along the free surface.  The
	adjoint boundary conditions are obtained by acting
	with the operator {\sf ii} on the redefined in
	$\ell$ 76 inner product {\sf iadj} ($\ell$ 84--86).

	\item Finally, the undetermined coefficients $f_i$
	are determined in $\ell$ 88--95 and the final output
	is written in the separate file ($\ell$ 97--100).
    \end{itemize}
\end{enumerate}

\bibliographystyle{plain}\bibliography{bib,ajr,new}
\end{document}